\documentclass[epj]{webofc}
\usepackage[varg]{txfonts}   
\usepackage[utf8]{inputenc}
\usepackage{xcolor}
\usepackage[compact]{titlesec}
\wocname{EPJ Web of Conferences}
\woctitle{ICNFP 2017}

\begin{document}
\selectlanguage{english}
\title{Threshold effects in hadron spectrum: a new spectroscopy?}

\author{P.~G.~Ortega\inst{1}\fnsep\thanks{\email{pgortega@usal.es}} \and
	J.~Segovia\inst{2} \and
        D.~R.~Entem\inst{1} \and
        F.~Fern\'andez\inst{1}
}

\institute{Grupo de F\'isica Nuclear and Instituto Universitario de F\'isica 
Fundamental y Matem\'aticas (IUFFyM), Universidad de Salamanca, E-37008 
Salamanca, Spain 
\and
Institut de F\'isica d'Altes Energies (IFAE) and Barcelona Institute of Science and Technology (BIST),
Universitat Aut\`onoma de Barcelona, E-08193 Bellaterra (Barcelona), Spain
}

\abstract{%
The exploration of energies above the open-flavor threshold in the meson spectra
has led to the appearance of unexpected states
difficult to accommodate in the naive picture of a bound state of a quark and an antiquark.
Many of such states are located close to meson-meson thresholds, which suggests 
that molecular structures may be a relevant component in the total wave function of such resonances.

In this work, the state of meson-meson molecules calculations is reviewed, 
using a non-relativistic constituent quark model that has been applied to a wide range of
hadronic observables, and therefore all model parameters are completely constrained. 
The model has been able to reproduce, among others, the properties of the $X(3872)$, described as a mixture of
$c\bar c$ and $D\bar D^\ast$ states, or the spectrum of the P-wave charm-strange mesons, which
are well reproduced only if $DK$ and $D^\ast K$ structures are taken into account. 

We show that such constituent quark model, which is able to describe the
ordinary heavy meson spectra, is also capable of providing a good description of many new states recently
reported.
}
\maketitle
\section{Introduction}
\label{intro}

Between November 1974, when the $J/\psi$ particle was discovered, until 2003, when the $X(3872)$ resonance was observed
by the Belle Collaboration, the quarkonium spectrum was satisfactorily
described by naive quark models. However, the exploration of energies above the open-flavor
threshold has led to the appearance of unexpected states
difficult to accommodate in the naive picture of a bound state of a quark and an antiquark.

The mentioned $X(3872)$ has properties that cannot be explained assuming a simple $q\bar q$ structure, as its decay into
$J/\Psi \pi \pi$ through a $\rho$ meson, which is an isospin violating decay. However
this property can be easily understood in a molecular picture in which the state is a 
$D\bar D^*$ bound state, due to the isospin violation in the $D$ and $D^*$ masses and the
close position of the state to the $D^0\bar D^{*\,0}$ threshold. 
In addition to the $X(3872)$, there are other states such as the $D_{s0}^\ast(2317)$ and $D_{s1}(2460)$, which can be described as a $DK$ and $D^\ast K$ molecules, respectively, 
or the $P^\ast_c(4380)$ and $P^\ast_c(4450)$, two exotic structures discovered in 2015 in the LHCb Collaboration in the $J/\psi p$ channel 
which lie near the $\bar D\Sigma^{(\ast)}_c$ thresholds and are compatible with pentaquark states.

In fact, the common feature among all these exotic states is the presence of nearby hadron-hadron thresholds. 
This coincidence reinforces the intuition that the residual
interaction between the two particles can bind them forming a molecular state. Then, one can
expect that near each meson-meson or baryon-meson thresholds molecular structures may appear. The description of this type of structures
is very tricky from a theoretical point of view because the strength of the residual interaction
is usually model dependent and one can generate spurious states if this interaction is not under control.

In this work we review the results of a widely-used Constituent Quark Model (CQM)~\cite{Vijande:2004he,Segovia:2008zza,Segovia:2016xqb}, which satisfactorily 
describes the meson spectra involving one or two heavy quarks, incorporating the effects of closeby meson-meson thresholds in a non-perturbative way.
This extension allowed to describe the inner structure of states with difficult assignment in the heavy quarkonium $Q\bar Q$ and $Q\bar q$ spectrum~\footnote{In this work 
$Q$ will stand for the $c$ and $b$ heavy quarks, while $q$ will refer to the light quarks $\{u,d,s\}$}.

Besides the meson spectra, the aforementioned CQM model is able to describe the baryonic phenomenology~\cite{Fernandez:1992xs,Valcarce:2005em,Ortega:2011zza} and, at the same time, 
some of the exotic states discovered in the baryon spectrum, such as the $P^\ast_c(4380)$ and $P^\ast_c(4450)$
pentaquarks as $\bar D \Sigma^\ast_c$ and $\bar D^\ast \Sigma_c$ molecules~\cite{Ortega:2016syt}, the $\Lambda_c(2940)^+$ as $D^\ast N$ bound state~\cite{Ortega:2012cx} 
or the $X_c(3250)$ one as a $D^\ast\Delta$ molecule~\cite{Ortega:2014fha}. 
Details for such calculations are out of scope for this review and the reader is kindly referred to the previous studies for further information.

\section{The model}
\label{sec:model}

\subsection{Constituent quark model}
\label{subsec:CQM}

Quark models have been a conceptually simple but powerful tool in describing the naive hadronic spectrum. 
Usually, in such models, quarks are massive fundamental objects and hadrons emerge from the interaction among quarks 
mediated by color interactions or bosons such as the pion, modeled from Quantum Chromodynamics (QCD).

The constituent quark model employed in the present work has been extensively described in the literature~\cite{Vijande:2004he},
but for the sake of clarity we will here review its most relevant aspects.
The original QCD Lagrangian satisfies the chiral symmetry. However, this symmetry appears
spontaneously broken in nature and, 
as a consequence, light quarks acquire a
dynamical mass. 
If we impose a Lagrangian which is invariant under chiral rotations, chiral fields must be included. Thus, 
the simplest Lagrangian with the latter properties can be expressed as 

\begin{equation}
\label{lagrangian}
{\mathcal L}
=\overline{\psi }(i\, {\slash\!\!\! \partial} -M(p^{2})U^{\gamma_{5}})\,\psi 
\end{equation}
where  $U^{\gamma_5}=e^{i\frac{
\lambda _{a}}{f_{\pi }}\phi ^{a}\gamma _{5}}$ is 
the Goldstone boson fields matrix and $M(p^2)$ the dynamical
constituent mass. 

The Goldstone boson field matrix $U^{\gamma_{5}}$ 
can be expanded in terms of boson fields,
\begin{equation}
U^{\gamma_{5}}=1+\frac{i}{f_{\pi }}\gamma^{5}\lambda^{a}\pi^{a}-\frac{1}{%
2f_{\pi}^{2}}\pi^{a}\pi^{a}+\ldots,
\end{equation}
where the first term of the expansion generates the constituent quark mass and the
second gives rise to a one-boson exchange interaction between quarks. The
main contribution of the third term comes from the two-pion exchange which
has been simulated by means of a scalar exchange potential.

Beyond the chiral symmetry breaking scale quarks still interact
through gluon exchanges described by the Lagrangian
\begin{equation}
\label{Lg}
{\mathcal L}_{gqq}=
i\sqrt{4\pi \alpha _{s} }\,\,\overline{\psi }\gamma _{\mu }G^{\mu
}_c \lambda _{c}\psi  \, ,
\end{equation}
where $\lambda_{c}$ are the SU(3) color generators and $G^{\mu}_c$ the
gluon field. 
The model is completed with the main QCD nonperturbative effect, confinement.
Such a term can be physically interpreted in a picture in which
the quark and the antiquark are linked by a one-dimensional color flux-tube.
The spontaneous creation of light-quark pairs may
give rise at some scale to a breakup of the color flux-tube~\cite{Bali:2005fu}. 

This interaction can be modeled with a screened potential~\cite{PhysRevD.40.1653} that takes
into account the breakup of the color flux-tube~\cite{Bali:2005fu} via the spontaneous creation of 
light-quark pairs, in such a way that the potential saturates at some interquark distance
\begin{equation}
V_{CON}(\vec{r}_{ij})=\{-a_{c}\,(1-e^{-\mu_c\,r_{ij}})+ \Delta\}(\vec{%
\lambda^c}_{i}\cdot \vec{ \lambda^c}_{j}).
\end{equation}
Explicit expressions for these interactions are given in Ref.~\cite{Vijande:2004he}.

\subsection{Coupled-channels formalism}
\label{subsec:coupled}

In this section we review the formalism for coupled channels, which describes the 
coupling of molecular structures with the heavy quarkonium $Q\bar Q$ spectrum 
(see Ref.~\cite{Ortega:2012rs} for further details).
By studying such interplay, the influence of the $Q \bar Q$ states on the dynamics
of the two meson states is studied, which allows to generate new states through 
the meson-meson interaction due to the coupling with heavy quarkonium states and to the underlying
$q-q$ interaction.

We start defining the global hadronic state as a combination of heavy quarkonium
and meson-meson channels, that is,
\begin{equation} 
\label{ec:funonda}
 | \Psi \rangle = \sum_\alpha c_\alpha | \psi_\alpha \rangle
 + \sum_\beta \chi_\beta(P) |\phi_{M_1} \phi_{M_2} \beta \rangle
\end{equation}
where $|\psi_\alpha\rangle$ are the $Q\bar Q$ eigenstates of the two body
Hamiltonian and 
$\phi_{M_i}$ are $Q\bar q$ (or $\bar Q q$) eigenstates describing 
the $M_i$ mesons. Moreover, 
$|\phi_{M_1} \phi_{M_2} \beta \rangle$ is the two meson state with $\beta$ quantum
numbers coupled to total $J^{PC}$ quantum numbers
and $\chi_\beta(P)$ is the relative wave 
function between the two mesons ${M_1}{M_2}$ in the molecule. 

The meson-meson interaction from the inner $q-q$ interactions is obtained by using the 
Resonating Group Method (RGM). Such method requires to know in advance the wave functions
of the mesons, which are calculated by solving the Sch\"odinger equation for the quark-antiquark
bound state using the Gaussian Expansion Method (GEM).
Following Ref.~\cite{Hiyama:2003cu} we employ Gaussian trial functions whose 
ranges are in geometric progression, which optimized the ranges with few free parameters. 
That way we obtain a dense distribution of Gaussians at small range, which is suitable
for short range potentials.

The coupling between the $Q\bar Q$ and the $Q\bar q-q\bar Q$ channels requires the creation 
of a light quark pair $q\bar q$. It is reasonable to assume that the latter process
will be described by a similar mechanism that describes the spectrum of the heavy quarkonium. 
However Ackleh {\it et al.}~\cite{PhysRevD.54.6811} showed that the quark pair creation $^3P_0$ 
model~\cite{Micu:1968mk} gives equivalent results to microscopic calculations,
while significantly simplifying the calculations. 
The non-relativistic reduction of the pair creation Hamiltonian is equivalent to the following transition
operator~\cite{Bonnaz:1999zj}
\begin{equation}
\mathcal{T}=-3\sqrt{2}\gamma'\sum_\mu \int d^3 p d^3p' \,\delta^{(3)}(p+p')
\times \left[ \mathcal Y_1\left(\frac{p-p'}{2}\right) b_\mu^\dagger(p)
d_\nu^\dagger(p') \right]^{C=1,I=0,S=1,J=0}
\label{TBon}
\end{equation}
where $\mu$ ($\nu=\bar \mu$) are the quark (antiquark) quantum numbers and
$\gamma'=2^{5/2} \pi^{1/2}\gamma$ with $\gamma= \frac{g}{2m}$ is a 
dimensionless constant that controls the strength of 
the $q\bar q$ pair creation from the vacuum.
From this operator, the transition potential $h_{\beta \alpha}(P)$ within 
the $^3 P_0$ model can be defined as
\begin{equation}
\label{Vab}
        \langle \phi_{M_1} \phi_{M_2} \beta | \mathcal{T}| \psi_\alpha \rangle =
        P \, h_{\beta \alpha}(P) \,\delta^{(3)}(\vec P_{\mbox{cm}}),
\end{equation}
being $P$ the relative momentum of the two meson state.

Adding the coupling with the heavy quarkonium states we end-up with a coupled-channel 
system for the quarkonium eigenstates and the meson-meson channels. 
Such equations can be simplified and expressed as an Schr\"odinger-type
equation~\cite{Ortega:2010qq} where, besides the potential coming from quark-quark interaction described by
the CQM, an effective potential emerges from the coupling with $Q\bar Q$ states,
\begin{equation}
V^{\rm eff}_{\beta'\beta}(P',P;E)=\sum_{\alpha}\frac{h_{\beta'\alpha}(P')
h_{\alpha\beta}(P)}{E-M_{\alpha}}.
\end {equation}

In order to describe resonances above thresholds within the same formalism, all the potentials must be analytically 
continued into the complex plane. Then, a more adequate method is 
obtained by solving the equations through the $T(\vec p,\vec p',E)$ matrix, solution of 
the Lippmann-Schwinger equation. 
Using the total potential $V_T(P',P;E)=V(P',P)+V^{\rm eff}(P',P;E)$, where $V(P',P)$ is the potential from CQM, 
the $T$-matrix can be factorized as 
\begin{equation}
\label{ec:tmat1}
T^{\beta'\beta}(P',P;E)=T^{\beta'\beta}_V(P',P;E)
+\sum_{\alpha,\alpha'}\phi^{\beta'\alpha'}(P';E)
\Delta_{\alpha'\alpha}^{-1}(E)\phi^{\alpha\beta}(P;E),
\end{equation}
where $T^{\beta'\beta}_V(P',P;E)$ the $T$ matrix of the CQM potential excluding the 
coupling to the $Q\bar Q$ pairs and $\Delta_{\alpha'\alpha}(E)$
is the propagator of the mixed state, 

\begin{equation}
 \Delta_{\alpha'\alpha}(E)=\left((E-M_{\alpha})\delta^{\alpha'\alpha}+\mathcal{
G}^{\alpha'\alpha}(E)\right),
\end{equation}
where $M_\alpha$ is the bare mass of the 
heavy quarkonium and $\mathcal{G}^{\alpha'\alpha}(E)$ the complete mass-shift. 
The $\phi^{\beta\alpha}(P;E)$ functions can be interpreted as the 
$^3P_0$ vertex dressed by the RGM meson-meson interaction.

Finally, resonances will emerge as poles of the $T$-matrix in the complex plane, namely as zeros of the 
determinant of the propagator of the mixed state, that is,
$\left|\Delta_{\alpha'\alpha}(\bar{E})\right|=0$, with $\bar{E}$ the pole position.

\section{Results}
\label{sec:results}

\subsection{Coupled-channels effects in the 3.9-GeV/$c^2$ charmonium spectra}
\label{subsec:charmonio}

The positive parity charmonium spectra, in particular $\chi_{cJ}(2P)$ at energies around 3.9 GeV/$c^2$,
is a very intriguing region after the discovery of several unexpected states that do not fit into the predictions of naive quark models.
The most representative resonance is the $X(3872)$, discovered in 2003 by the Belle Collaboration~\cite{Choi:2003ue}. 
This state, with $J^{PC}=1^{++}$~\cite{PhysRevD.92.011102}, decays through the $J/\psi\rho$ and $J/\psi\omega$ channels which are, respectively, forbidden and OZI-suppressed for a $c\bar c$ configuration.
Besides, many other structures have been described over the years in this energy region, such as the $Y(3940)$ resonance, discovered by Belle~\cite{PhysRevLett.94.182002}, the $X(3940)$ one~\cite{PhysRevLett.98.082001} or the $X(3930)$ state, a $J^{PC}=2^{++}$ peak measured in 2006 by the same Collaboration in the mass spectrum of the $D\bar D$ mesons produced by $\gamma \gamma$ fusion, with $M=3929\pm 6$ MeV$/c^2$ and $\Gamma=29\pm 10$ MeV, which was rapidly assigned to the $\chi_{c2}(2P)$ state~\cite{Olive:2016xmw}. 

Furthermore, a recent controversy has appeared about the nature of the $X(3915)$ resonance~\cite{Uehara:2009tx,Lees:2012xs}, discovered in two-gamma fusion processes.
This resonance has been traditionally assigned to a $J^{PC}=0^{++}$ state, but recent works favor a $J^{PC}=2^{++}$ nature, implying a large
non-$q\bar q$ component for the $X(3915)$ state~\cite{Guo:2012tv,Olsen:2014maa,Zhou:2015uva}.
Moreover, a new charmonium-like state decaying to $D\bar D$, called $X(3860)$, has been reported by the Belle Collaboration~\cite{Chilikin:2017evr}, having a mass of $3862_{-32\,-13}^{+26\,+40}$ MeV/$c^2$ and a width of $201_{-67\,-82}^{+154\,+88}$ MeV. 
For the latter, the $J^{PC}=0^{++}$ option is favored over the $2^{++}$ hypothesis, but its quantum numbers are not definitively determined. 

Such experimental situation requires more theoretical efforts in order to unveil the nature of these resonances.
Remarking that the $X(3872)$, $X(3940)$, $X(3915)$, $X(3860)$, $X(3930)$ and $Y(3940)$ resonances all belong to the same energy region it is expected that 
these states are highly influenced by the interplay between two and four quark channels.
For that reason, based on the initial analysis of the $J^{PC}=1^{++}$ sector by Refs.~\cite{Ortega:2010qq,Ortega:2012rs}, 
a coupled-channels calculation for the $0^{++}$ and $2^{++}$ channels is performed~\cite{Ortega:2017qmg}, including the $c\bar c\,(2^3P_J)$ bare charmonium states, predicted
by the CQM described above, which are coupled to the closest meson-meson thresholds. 

In principle, we would need to couple with the infinite number of meson-meson thresholds,
but it has been argued by many theorists~\cite{Swanson:2005rc, Barnes:2007xu} 
that the only relevant thresholds are those close to the naive states, having the 
rest a little effect which can be absorbed in our quark model parameters.
For the present calculation, the $c\bar c\,(2^3P_J)$ bare meson states will be coupled with the following meson-meson channels, where the threshold energies are indicated in parenthesis:
\begin{itemize}
 \item $0^{++}:$ $D\bar D$ (3734 MeV/c$^2$), $\omega J/\psi$ (3880 MeV/c$^2$), $D_s\bar D_s$ (3937 MeV/c$^2$) and $D^\ast\bar D^\ast$ (4017 MeV/c$^2$),
 \item $2^{++}:$ $D\bar D$ (3734 MeV/c$^2$), $\omega J/\psi$ (3880 MeV/c$^2$), $D_s\bar D_s$ (3937 MeV/c$^2$),$D\bar D^\ast+h.c.$ (3877 MeV/c$^2$) and $D^\ast\bar D^\ast$ (4017 MeV/c$^2$),
\end{itemize}
where the $D^\ast D^\ast$ channel is included because it is the only threshold which couples to the $c\bar c\,(2^3P_0)$ in $S$ wave and, 
hence, its effect is expected to be essential to describe the $0^{++}$ sector. 
Using the original parameters of Ref.~\cite{Ortega:2010qq} (which will be denoted as {\it model A}) we obtain the masses and widths shown in Table~\ref{tab:model A}.
To explore the robustness of the results, in addition to the results for model A, a second calculation (named {\it model B}) has been performed, where the bare mass of 
the $c\bar c\,(2^3P_J)$ pairs has been slightly changed ($0.25\%$) and we have taken the value of the $^3P_0$ strength parameter $\gamma$ from Ref.~\cite{Segovia:2012cd}. 
Such variations are within the uncertainties of the model parameters.

\begin{table*}[!t]
\caption{\label{tab:model A} Mass and decay width, in MeV, and probabilities of 
the different channels, for models A and B.}
\begin{tabular}{ccccccccc}
\hline\hline
$J^{PC}$ & Mass & Width & ${\cal P} [c\bar{c}] $ & ${\cal P}[D\bar D]$ & ${\cal P}[D \bar D^{\ast}]$ & ${\cal 
P}[\omega J/\psi]$ & ${\cal P}[D_s\bar D_s]$ & ${\cal P}[D^{\ast} \bar D^{\ast}]$\\
\hline
\multicolumn{9}{c}{Model A}\\
\hline
$0^{++}$ & $3890.3$ & $6.7$ & $44.1\%$ & $21.6\%$ & $-$ & $28.4\%$ & $2.6\%$ & $3.3\%$ \\
$0^{++}$ & $3927.4$ & $229.8$ & $19.2\%$ & $66.3\%$ & $-$ & $5.3\%$ & $3.7\%$ & $5.5\%$ \\
$2^{++}$ & $3925.6$ & $19.0$ & $42.2\%$ & $11.3\%$ & $37.0\%$ & $4.0\%$ & $0.4\%$ & $5.1\%$ \\
\hline
\multicolumn{9}{c}{Model B}\\
\hline
$0^{++}$ & $3889.0$ & $11.8$ & $43.5\%$ & $27.3\%$ & $-$ & $20.4\%$ & $3.8\%$ & $4.9\%$ \\
$0^{++}$ & $3947.5$ & $201.6$ & $19.4\%$ & $66.0\%$ & $-$ & $3.7\%$ & $8.0\%$ & $2.9\%$ \\
$2^{++}$ & $3915.1$ & $19.8$ & $37.8\%$ & $14.1\%$ & $36.4\%$ & $5.12\%$ & $0.4\%$ & $6.1\%$ \\
\hline\hline
\end{tabular}
\end{table*}

Both models predict the same number of states: two resonances with $J^{PC}\!\!=\!0^{++}$, a broad one and a narrow one, and only 
one with $J^{PC}\!\!=\!2^{++}$. The $J^{PC}\!\!=\!2^{++}$ state is compatible with the mass and width of the $X(3930)$ one for model A, but its mass
lowers to the $X(3915)$ region for model B. 

In the $J^{PC}\!\!=\!0^{++}$ region both models describe the same scenario. On the one hand, the mass of the narrow state is closer to the new $X(3860)$ resonance than to the $X(3915)$. However, the predicted width is smaller than the experimental one, which could be connected with the position of the node in the $c\bar c\,(2^3P_0)$ bare wave function, which affects the $^3P_0$ transition 
amplitudes and, hence, causes a higher sensitivity of the width to small changes in the wave function structure or, alternatively, the mass of the $X(3860)$ resonance. 
The mass of the second $J^{PC}\!\!=\!0^{++}$ state suggests a preliminary assignment to the $Y(3940)$ resonance, but, as before, the value of the width is far from the experimental value. 
Alternatively, one can identify the second $0^{++}$ state with the $X(3860)$, as it is a broad state whose width matches with the experimental data and its mass is close to the experiments, which value reaches more than $3900$ MeV. In this picture, the first state is too narrow and can hardly be observed in the experiment of Ref.~\cite{Chilikin:2017evr}.
We remark that none of the $J^{PC}\!\!=\!0^{++}$ states seems to favour a $X(3915)$ assignment. 

In contrast, the nature of the predicted $J^P=2^+$ state is compatible with both $X(3915)$ and $X(3930)$, within the uncertainties of the model. 
In order to further analyze the properties of the $X(3915)$ resonance, we can explore its decays. If we assume the $X(3915)$ is one of the two $J^{PC}\!\!=\!0^{++}$ states we are unable to describe the experimental results. Instead, assuming the $X(3915)$ and $X(3930)$ are the same $J^{PC}\!\!=\!2^{++}$ resonance, we found the results quoted in Table~\ref{tab:BR J2},
where we also include the decay to the $D\bar D^*$ channel.
The results for both model A and B are very similar and not far from the experimental data. Then, the results suggest that the $X(3915)/X(3930)$ resonances are $J^{PC}\!\!=\!2^{++}$, conclusion that agrees with Ref.~\cite{Branz:2010rj}.

\begin{table*}[!t]
\caption{\label{tab:BR J2} Product of the two-photon decay width and the branching fraction to different channels (in eV) for the 
$J^{PC}=2^{++}$ sector for each model, and comparison with Belle and BaBar Collaboration experimental results.}
\begin{tabular}{ccccc}
\hline\hline
 & Belle & BaBar & model A &  model B  \\
\hline
$\Gamma_{\gamma \gamma} \times {\cal B}(2^{++}\to \omega J/\psi )$ & $18\pm 5 \pm 2$~\cite{Uehara:2009tx} & $10.5\pm1.9\pm 0.6$~\cite{Lees:2012xs} & $20.9$ & $24.9$  \\
$\Gamma_{\gamma \gamma} \times {\cal B}(2^{++}\to D\bar D )$ & $180\pm 50\pm 30 $~\cite{Uehara:2005qd} & $249\pm 50\pm 40$~\cite{Aubert:2010ab} & $75.4$ & $81.4$  \\
$\Gamma_{\gamma \gamma} \times {\cal B}(2^{++}\to D\bar{D^\ast})$ & - & - & $196.0$ & $151.9$  \\
\hline\hline
\end{tabular}
\end{table*}

\subsection{$X(4140)$, $X(4274)$, $X(4500)$ and $X(4700)$ states}
\label{subsec:charmonio}

Recently, the LHC Collaboration measured four structures in the $B^{+}\to J/\psi\phi K^{+}$
decay~\cite{Aaij:2016iza, Aaij:2016nsc}, the
$X(4140)$, $X(4274)$, $X(4500)$ and $X(4700)$ resonances. The only one that had already been measured at that time
was the $X(4140)$ state, previously seen by CDF~\cite{Aaltonen:2011at}, D0~\cite{Abazov:2015sxa}, 
CMS~\cite{Chatrchyan:2013dma}, Belle~\cite{Shen:2009vs} and 
BaBar~\cite{Lees:2014lra} Collaborations.
Their quantum numbers  were determined to be $J^{PC}=1^{++}$ for the $X(4140)$ and $X(4274)$, and $J^{PC}=0^{++}$
for the $X(4500)$ and $X(4700)$ resonances. 

From a theoretical point of view, many models have been proposed to describe the $X(4140)$ resonance, which includes 
tetraquark~\cite{Lebed:2016yvr} and molecule~\cite{Zhang:2009st} interpretations. 
However, few studies assigned it to a $1^{++}$ state as discovered by LHCb.
Therefore, the interest on these states resides in the difficulty to describe their structure as molecules or tetraquarks.

In order to explore the structure of such states with our coupled-channels approach we performed a
simultaneous analysis of the $0^{++}$ and $1^{++}$ channels in the energy region of those resonances~\cite{Ortega:2016hde}, 
which is at least $200$ MeV above the energy of the calculations performed in the previous section,
thus we assume they can be calculated independently.
For this energy region, then, the nearby thresholds are $D_{s}D_{s}^{\ast}$, $D_{s}^{\ast}D_{s}^{\ast}$ 
and $J/\psi\phi$ for $J^{PC}=1^{++}$, including as well
the closest naive $c\bar c$ states, that is, the $c\bar c\,(n^3P_1)$; and
$D^{\ast}D_{1}^{(\prime)}$, $D_{s}D_{s}$, $D_{s}^{\ast}D_{s}^{\ast}$ and 
$J/\psi\phi$ for $J^{PC}=0^{++}$, coupled to $c\bar c\,(n^3P_0)$, with $n=3,4,5$.

\begin{table}[!t]
\begin{center}
\caption{\label{tab:r1} Mass, in MeV, total decay width, in MeV, and 
probability of each Fock component, in \%, for the $X(4500)$ and $X(4700)$ 
mesons. The calculated widths include both the contributions of the $c\bar c$ 
and molecular components. The results have been calculated in the 
coupled-channel quark model.}
\begin{tabular}{cccccccc}
\hline
\hline
Mass & Width & ${\cal P}_{c\bar c}$ & ${\cal P}_{D^{\ast}D_{1}}$ & ${\cal 
P}_{D^{\ast}D_{1}^{\prime}}$ & ${\cal P}_{D_{s}D_{s}}$ & ${\cal 
P}_{D_{s}^{\ast}D_{s}^{\ast}}$ & ${\cal P}_{J/\psi\phi}$ \\
\hline
$4493.6$ & $79.2$ & $57.2$ & $8.4$ & $33.1$ & $0.9$  & $0.4$ & $<0.1$ \\
$4674.1$ & $50.2$ & $47.6$ & $27.2$ & $21.0$ & $1.6$ & $2.6$ & $<0.1$ \\
\hline
\hline
\end{tabular}
\end{center}
\end{table}

\begin{table}[!t]
\begin{center}
\caption{\label{tab:r3} Same caption as Table~\ref{tab:r1} for the $X(4274)$ meson.}
\begin{tabular}{cccccc}
\hline
\hline
Mass & Width & ${\cal P}_{c\bar c}$ & ${\cal P}_{D_{s}D_{s}^{\ast}}$ & ${\cal 
P}_{D_{s}^{\ast}D_{s}^{\ast}}$ & ${\cal P}_{J/\psi\phi}$ \\
\hline
$4242.4$ & $25.9$ & $48.7$ & $43.5$ & $5.0$ & $2.7$ \\
\hline
\hline
\end{tabular}
\end{center}
\end{table}

For the $J^{PC}=0^{++}$ (see Table~\ref{tab:r1}), the masses of the states are close to those associated with the 
bare $q\bar q$ $J^{PC}=0^{++}$ $4P$ and $5P$ states. 
The resonance with a mass of $4493.6\,{\rm MeV}$ is almost a pure $4P$ $c\bar 
c$ state whereas the one with a mass of $4674.1\,{\rm MeV}$ is an almost-pure 
$5P$ $c\bar c$ state. The masses and widths are compatible with those reported by the LHCb.
For the $J^{PC}=1^{++}$ channel (see Table~\ref{tab:r3}), only one state with mass $4242.4\,{\rm MeV}$ is found. 
This state has $48.7\%$ of the $3P$ charmonium state 
and around $43.5\%$ of $D_{s}D_{s}^{\ast}$ component. 
Our total decay width is still compatible with the LHCb 
result, but indicates that the lower CDF and CMS measurements are in better 
agreement with our prediction.

\begin{figure}[!t]
\begin{center}
\sidecaption
\includegraphics[width=0.5\textwidth]{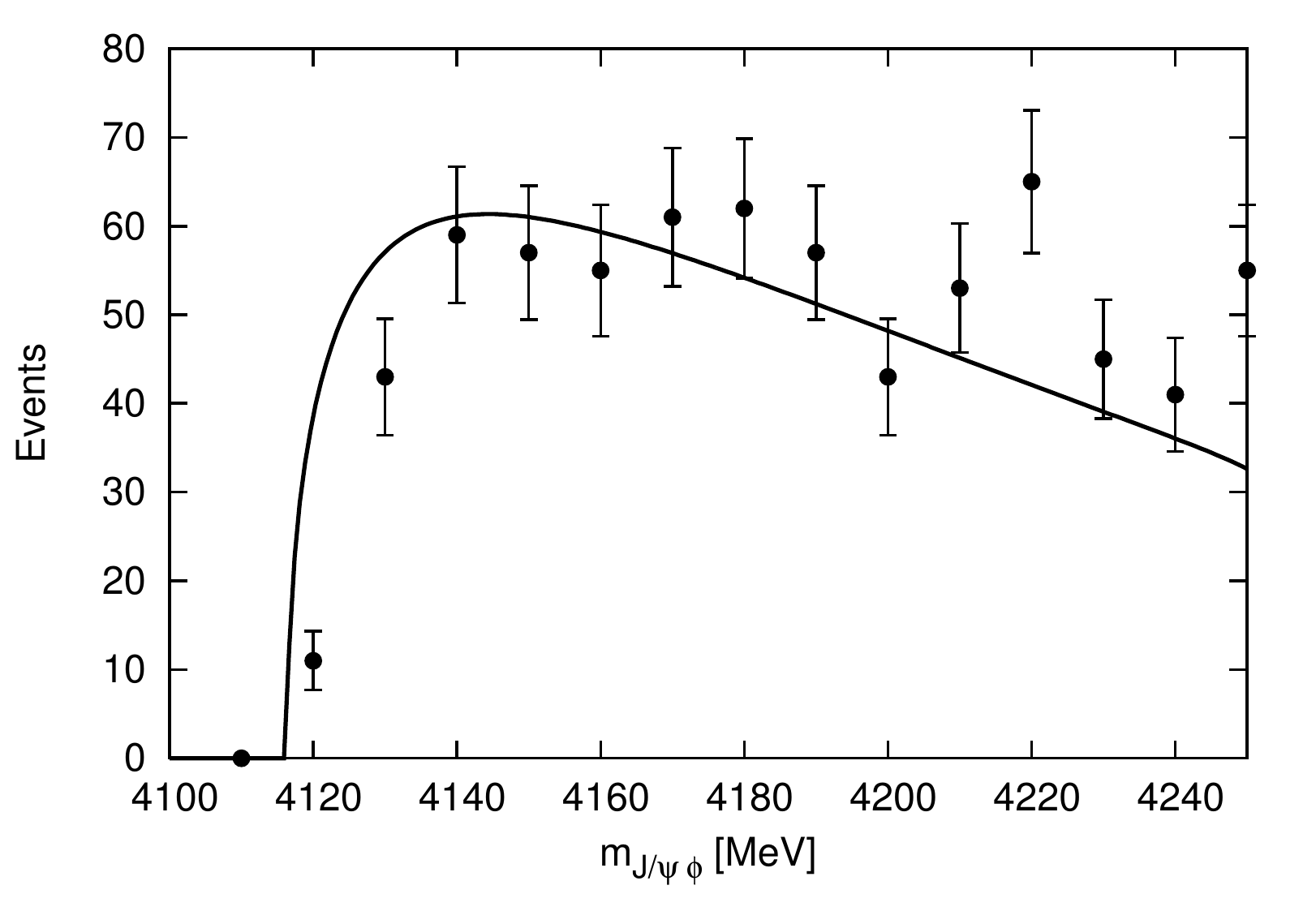}
\caption{\label{fig:f2} Line-shape prediction of the $J/\psi\phi$ channel. The 
solid curve shows the full result including both production mechanism: the direct 
generation of $J/\psi$ and $\phi$ mesons from a point-like source and 
the production via intermediate $c\bar c$ states.}
\end{center}
\end{figure}

Finally, no signal for the $X(4140)$ state is found, neither bound nor virtual.
Hence, we analyze the line shape of the $J/\psi\phi$ channel as an attempt to explain 
the $X(4140)$ as a simple threshold cusp. 
We evaluate the production of $J/\psi\phi$ pairs via two main mechanisms: (i) 
the direct generation of $J/\psi$ and $\phi$ mesons from a point-like source 
and (ii) the production via intermediate $c\bar c$ states (see Ref.~\cite{Ortega:2016hde}
for details). 
Figure~\ref{fig:f2} compares our result with that reported by the LHCb 
Collaboration in the $B^+\to J/\psi\phi K^+$ decays. The rapid increasing 
observed in the data near the $J/\psi\phi$ threshold corresponds with a bump in 
the theoretical result just above such threshold. This cusp is too wide to be 
produced by a bound or virtual state below the $J/\psi\phi$ threshold. We
find that the production via intermediate $c\bar c$ states is dominant at low values of the 
invariant mass, which is reasonable as the point-like production should be 
suppressed.

\subsection{Heavy-strange low-lying spectrum}
\label{subsec:DsBs}

Mesons containing a heavy ($Q=c,b$) quark and a light antiquark are interesting systems. Such quark-antiquark combination has great advantages from a
theoretical point of view, as heavy quark spin symmetry (HQSS)~\cite{Isgur:1989vq} is approximately fulfilled by QCD and the parity and
total angular momentum of the light antiquark ($j_{\bar q}$) can be assumed as conserved quantities.
This implies that the states can be arranged in doublets labelled by their parity and $j_{\bar q}$, and members within a doublet
are described by the same dynamics and, thus, are degenerated in mass (up to $\Lambda_{\rm QCD}/m_Q$ corrections).

The interest on heavy-strange spectrum ($Q\bar s$) gain attention in 2003, with the discovery of the 
$D_{s0}^{\ast}(2317)$ ($J^P=0^+$) and $D_{s1}(2460)$ ($J^P=1^+$) states by the BaBar~\cite{Aubert:2003fg}
and Cleo
collaborations~\cite{Besson:2003cp} in  the $D_s^{(\ast)+}\pi^0$
invariant mass spectrum.
Contrary to quark model and lattice QCD (LQCD) predictions~\cite{DiPierro:2001dwf, Dougall:2003hv}, the states lie below $D^{(\ast)}K$ thresholds, so their masses and widths were lower than expected.
Such states are members of the $j_{\bar q}^P=\tfrac{1}{2}^+$ doublet, which  
strongly couples to S-wave $D^{(\ast)}K$ pairs, so the influence of such threshold may be relevant to describe the resonances.

In this section we will show the results of calculations for the low-lying charm-strange~\cite{Ortega:2016mms} and bottom-strange~\cite{Ortega:2016pgg} spectrum including the $Q\bar s(1^3P_0)$ CQM state plus the $H_QK$~\footnote{For now on, we will denote $H_Q^{(\ast)}$ as the $D^{(\ast)}$ and $B^{(\ast)}$ heavy mesons, depending on the system under study.} channel
for the $J^P=0^+$ sector and the $Q\bar s(1^1P_1)+Q\bar s(1^3P_1)$ bare meson states coupled to $H_Q^\ast K$ channel for the $J^P=1^+$ channel.
For this calculation we will add to the one-gluon exchange potential $(\alpha_{s})$ its one-loop corrections $(\alpha_{s}^{2})$, which has an impact in the
bare masses of the $J^P=0^+$ states, lowering their mass and allowing a major effect of the $H_QK$ threshold in the coupled-channels approach~\cite{Segovia:2012yh}.

Tables~\ref{tab:Ds0} and~\ref{tab:Bs0} show the masses, widths and probabilities of the $J^P=0^+$ low-lying $P$-wave 
charm-strange and bottom-strange mesons predicted by the coupled-channels calculation. For both sectors, $D_s$ and $B_s$, we find one bound state
below the lower $H_QK$ threshold, with a larger $Q\bar s$ component, which mimics the experimental situation. The coupling with the nearby threshold
is able to shift the mass of the bare $Q\bar s$ below the threshold, obtaining a narrow state. For the bottom-strange sector the situation is the same, but
a wide resonance above the $BK$ threshold emerges together with the bound state, which is in agreement with previous studies~\cite{Albaladejo:2016ztm}.

\begin{table}[!t]
\begin{center}
\caption{\label{tab:Ds0} Masses, widths (in MeV) and probabilities (in \%) of the different 
Fock components for the states found in the $J^P=0^{+}$ $c\bar{s}$ sector.}
\begin{tabular}{cccccc}
\hline\hline
State & Mass & Width & ${\cal P}[Q\bar{s}\,(^{3}P_{0})]$ & ${\cal P}[DK- S wave]$ \\
\hline
$D_{s0}^{\ast}(2317)$ & $2323.7$ & $0$ &  $66.3\%$ &  $33.7\%$ \\
\hline\hline
\end{tabular}

\end{center}
\end{table}

\begin{table}[!t]
\begin{center}
\caption{\label{tab:Bs0} Same caption as Table~\ref{tab:Ds0} for the $J=0^+$ bottom-strange sector.}
\begin{tabular}{cccccc}
\hline\hline
State & Mass & Width & ${\cal P}[Q\bar{s}\,(^{3}P_{0})]$ & ${\cal P}[BK- S wave]$ \\
\hline
$B_{s0}^{\ast}$ & $5741.4$ & $0$     & $61.2$ &  $38.8$ \\
$B_{s0}^R$       & $5877.9$ & $303.5$ & $83.7$ &  $16.3$ \\
\hline\hline
\end{tabular}

\end{center}
\end{table}

\begin{table*}[!t]
\caption{\label{tab:Ds2460P} Mass and decay width (in MeV), and probabilities (in \%) of 
the different Fock components of the ${D_{s1}(2460)}$ and 
$D_{s1}(2536)$ states. 
}
\begin{tabular}{ccccccc}
\hline
\hline
State & Mass & Width & ${\cal P}[q\bar{q}\,(^{1}P_{1})]$ & ${\cal 
P}[q\bar{q}\,(^{3}P_{1})]$ & ${\cal P}[D^{\ast}K(S-wave)]$ & ${\cal 
P}[D^{\ast}K(D-wave)]$\\
\hline
$D_{s1}(2460)$ & $2484.0$ & $0.00$ & $12.1\%$ & $33.6\%$ & $54.1\%$ & $0.2\%$ \\
$D_{s1}(2536)$ & $2535.2$ & $0.56$ & $31.9\%$ & $14.5\%$ & $16.8\%$ & $36.8\%$ \\
\hline
\hline
\end{tabular}
\end{table*}

\begin{table*}[!t]
\begin{center}
\caption{\label{tab:Bs1Pro} Same caption as Table~\ref{tab:Ds2460P} for the $J=1^+$ bottom-strange sector.}
\begin{tabular}{ccccccc}
\hline
\hline
State & Mass & Width & ${\cal P}[q\bar{q}\,(^{1}P_{1})]$ & ${\cal 
P}[q\bar{q}\,(^{3}P_{1})]$ & ${\cal P}[B^{\ast}K(S-wave)]$ & ${\cal P}[B^{\ast}K(D-wave)]$ \\
\hline
$B_{s1}^{\prime}$ & $5792.5$ & $0.000$ &  $13.2\%$ & $42.6\%$ & $44.2\%$ & $0.0\%$ \\
$B_{s1}(5830)$ & $5832.9$ & $0.058$ & $35.4\%$ &  $12.1\%$ & $15.9\%$ & $36.6\%$\\
$B_{s1}^R$ & $5940.4$ & $271.3$ & $21.4\%$ & $58.8\%$ & $19.8\%$ & $0.0\%$ \\
\hline
\hline
\end{tabular}

\end{center}
\end{table*}

The results for $J^P=1^+$ are shown in Table~\ref{tab:Ds2460P} for $c\bar s$ and Table~\ref{tab:Bs1Pro} for $b\bar s$ systems.
Again, for each sector, we find one bound state below the $H_Q^\ast K$. The charm-strange case is in agreement with
the mass of the $D_{s1}(2460)$, whereas the $B_{s1}^\prime$ mass is a prediction of our model. The $D_{s1}(2536)$ and $B_{s1}(5830)$
mesons are the members of the $j_{\bar q}^P=3/2^+$ HQSS doublet. It is relevant to realize that the $Q\bar s$ component in the wave 
function of the $D_{s1}(2536)$ and $B_{s1}(5830)$ mesons holds quite well the $^{1}P_{1}$ and 
$^{3}P_{1}$ composition predicted by HQSS. This is crucial in order to have a very narrow 
state and describe well its decay properties. 

We also notice that when we couple to the nearby threshold, the meson-meson component is $\sim50\%$ 
for the narrow states $D_{s1}(2460)$, $D_{s1}(2536)$, $B_{s1}^\prime$ and $B_{s1}(5830)$ mesons. 
For the $1^+$ sector this is in agreement with lattice calculations~\cite{Lang:2014yfa,Lang:2015hza}, which find similar overlaps of 
the physical states with the quark-antiquark and meson-meson interpolators, and, in the case of $c\bar s$, with
the one reported in Ref.~\cite{Torres:2014vna}. However, for the $0^+$ $c\bar s$ sector we predict a $\sim 34\%$ molecular component, 
whereas other studies describe a $\sim 70\%$ of molecule (see Ref.~\cite{Ortega:2016mms} for discussion).
Finally, for the bottom-strange sector we also find a wide state which does not appear in the charm-strange sector.

\section{Discussion}
\label{sec:discussion}

As a summary, we have analysed in a coupled-channel quark model the molecular
candidates in the $Q\bar Q$ and $Q\bar s$ systems, discussing the most relevant sectors. 
These results constitute a prominent example of the interplay 
between quark and meson degrees of freedom in near open-flavoured threshold 
regions. Depending on the dynamics of the system, the presence of these 
thresholds can generate new states, simply renormalizes the masses of the bare 
$q\bar q$ states or produces cusp effects at threshold when the interactions 
are not strong enough to produce bound states.

Finally, these results also reinforce the validity of the constituent quark 
model to qualitatively describe the phenomenology of the excited heavy quark 
meson states and get insights on the dynamics that leads their formation.

\section*{Acknowledgments}

This work has been partially funded by European Union's Horizon 2020 research and 
innovation programme under the Marie Sk\l{}odowska-Curie grant agreement no. 665919; by Ministerio de Ciencia y Tecnolog\'ia 
under Contract no. FPA2013-47443-C2-2-P; by Ministerio de Econom\'ia, Industria y 
Competitividad under Contract nos. FPA2016-77177-C2-2-P,  FPA2014-55613-P and SEV-2016-0588; 
and by Junta de Castilla y Le\'on and European Regional Development Funds (ERDF) under Contract no. SA041U16.

\bibliography{biblio}

\end{document}